\documentstyle[prl,aps,epsf,rotate]{revtex}

\begin{document}
\draft
\tighten
\title
{\LARGE Crossover from Self--similar to Self--affine Structures \\
in Percolation}
\vspace{1cm}

\author{Erwin Frey}
\address{Institut f\"ur Theoretische Physik,
Physik-Department der Technischen Universit\"at M\"unchen, \\
James-Franck-Stra\ss e, D-85747 Garching, Germany}
\author{Uwe Claus T\"auber}
\address{Lyman Laboratory of Physics, Harvard University,
Cambridge, Massachusetts 02138, U.S.A.}
\author{Franz Schwabl}
\address{Institut f\"ur Theoretische Physik,
Physik-Department der Technischen Universit\"at M\"unchen, \\
James-Franck-Stra\ss e, D-85747 Garching, Germany}
\date{\today}
\maketitle
\widetext
\centerline{(to be published in Europhys. Lett. 1994)}

\baselineskip = 20pt

\begin{abstract}

We study the crossover from self--similar scaling behavior to asymptotically
self--affine (anisotropic) structures. As an example, we consider bond
percolation with one preferred direction. Our theory is based on a
field--theoretical representation, and takes advantage of a renormalization
group approach designed for crossover phenomena. We calculate effective
exponents for the connectivity describing the entire crossover region from
isotropic to directed percolation, and predict at which scale of the anisotropy
the crossover should occur. We emphasize the broad range of applicability of
our method.
\end{abstract}

\pacs{PACS numbers: 0520, 0540, 6460}


\baselineskip = 20pt

Scale--invariance appears in a great variety of physical problems and processes
\cite{Man82}. The simplest scaling laws are {\it isotropic}, and thus describe
{\it self--similar} structures; on the other hand, in growth models frequently
{\it self--affine} clusters emerge, which we consider as defined by {\it
anisotropic} scaling. The important point now is that whether a structure
appears self--similar or self--affine is always a matter of scale, i.e.: a
structure, which on a large length scale is characterized by anisotropic
scaling laws, may be indistinguishable from a self--similar entity on
sufficiently small scales.

Perhaps the most simple growth models incorporating these features are provided
by percolation \cite{Ess80}. In ordinary, isotropic percolation sites or bonds
are filled at random with equal probability $p$, and the emerging clusters are
self--similar. However, in directed percolation \cite{Kin83}, there is one
distinct preferred direction (which we shall call the $t$ direction) with a
bias along the positive $t$ direction, and anisotropic scaling laws apply. If
the effective anisotropy parameter $g$ is low, one expects almost isotropic
scaling behavior in a large region of parameter space, and only if the
percolation threshold $p_c$ is approached the asymptotic self--affine scaling
will become apparent, if one views the system at very large length scales.

Our aim is to provide a quantitative description of the crossover from
isotropic to directed percolation, and to calculate the characteristic
anisotropy scale, at which the scaling behavior turns over from
self--similarity to self--affinity. For these issues, the central quantity of
interest is the pair correlation function $G({\bf r}_2,{\bf r}_1)$, which
measures the probability that sites ${\bf r}_2 = ({\bf x}_2,t_2)$ and
${\bf r}_1 = ({\bf x}_1,t_1)$ belong to the same cluster. Lines of constant $G$
describe the average shape of the percolating structure. Following the work of
Cardy et al. \cite{Car80,Ben84}, we apply their mapping of the statistical
percolation problem onto a field--theoretical representation. The final result
for $G^0$ (the superscript ``0'' denotes unrenormalized quantities) is a sum
$G^0({\bf r}_2;{\bf r}_1) = \sum_{m,n=1}^\infty [(-i)^{m+n-2} / m! n!]
G^0_{mn}({\bf r}_2;{\bf r}_1)$, where
\begin{equation}
     G^0_{mn} ({\bf r}_2;{\bf r}_1) = P \int {\cal D} [\phi_0,{\tilde \phi}_0]
           \phi_0({\bf r}_2)^m {\tilde \phi}_0({\bf r}_1)^n
           e^{- {\cal J} [\phi_0,{\tilde \phi}_0]} \; .
\label{1}
\end{equation}
$P$ is an operator projecting out those contributions which violate
``causality'', and the probability measure explicitly reads \cite{Car80,Ben84}
\begin{equation}
     {\cal J} = \! \int \! \! d^Dx \! \! \int \! \! dt
       \biggl\{ {\tilde \phi}_0  \!
         \left[ r_0 - {\bf \nabla}^2 - {1 \over c_0^2} \partial_t^2
           + {2 g_0 \over c_0} \partial_t \right] \! \phi_0
           + {u_0 \over 2} \left[ {\tilde \phi}_0^2 \phi_0
              - {\tilde \phi}_0 \phi_0^2 \right] \biggr\} \; ,
\label{2}
\end{equation}
with $r_0 \propto p - p_c$. The anisotropy of the underlying percolation
problem is reflected in the parameter $g_0$. For $g_0 = 0$ the problem is
symmetric with respect to inversion of the lattice ${\bf r} \rightarrow
-{\bf r}$, which corresponds to isotropic bond percolation \cite{Ben84}. For
any non--zero $g_0$ the inversion symmetry is broken in the $t$ direction. A
scaling anlaysis shows that $g_0$ grows under rescaling. For $g_0 \rightarrow
\infty$, with $c_0 \rightarrow \infty$ such that $g_0/c_0$ remains finite, one
arrives at Reggeon field theory describing directed bond percolation
\cite{Car80}. On the basis of Eqs.~(\ref{1},\ref{2}) the perturbation expansion
\cite{Ami84,Jan76} with respect to the nonlinearity $u_0$ may be constructed.

In order to derive the anomalous dimensions leading to non--Gaussian behaviour
we proceed to study the ultraviolet divergences at the upper critical dimension
$d_c$. In the dimensional regularization scheme \cite{Hoo72} these UV
singularities appear as poles $\propto 1/(d - d_c)$. The technical difficulty
here is now that these poles will be different in the two limiting cases
$g_0 = 0$ and $g_0 \rightarrow \infty$. In addition, via a simple rescaling of
the model one finds that the upper critical dimensions for isotropic (I) and
directed (D) percolation differ, namely $d_c^{\rm I} = 6$ and $d_c^{\rm D} =
5$, respectively. Thus the ``traditional'' procedure of investigating the
scaling behavior near one of the fixed points in the framework of a $(d_c - d)$
expansion, and then describing the crossover by calculating the accompanying
scaling function, is bound to fail.

However, a certain extension of Amit and Goldschmidt's {\it ``generalized
minimal subtraction'' procedure} \cite{Ami78} has proven to be most successful
in a variety of interesting crossover scenarios \cite{Law81,Fre88,Tau92}.
Indeed, by refraining from any $(d_c - d)$ expansion, we have demonstrated that
the crossover between fixed points of different upper critical dimension can be
consistently incorporated into the formalism \cite{Fre93}. The essential point
is that one has to assure that the renormalization constants take account of
the UV poles for {\it any} value of the anisotropy parameter $g_0$, {\it
including} the limit $g_0 \rightarrow \infty$. Thus the $Z$ factors become
functions of both the anharmonic coupling {\it and} the additional mass $g_0$.
By keeping the full dimension dependence of the corresponding residua, a smooth
interpolation between the different scaling regimes is achieved (for details we
refer to our forthcoming paper \cite{Fre93}). The prize that we have to pay is
that (i) there is no a-priori small expansion parameter, and (ii) the flow
equations require a numerical solution.

Thus we define renormalized fields $\phi = Z_\phi^{1/2} \phi_0$ and ${\tilde
\phi} = Z_\phi^{1/2} {\tilde \phi}_0$, and dimensionless renormalized
parameters $r = Z_\phi^{-1} Z_r (r_0 - r_{0c}) \mu^{-2}$, $c^2 = Z_\phi
Z_c^{-1} c_0^2$, $g = Z_\phi^{-1/2} Z_c^{-1/2} Z_g g_0 \mu^{-1}$, and $u =
Z_\phi^{-3/2} Z_u u_0 B_d^{1/2} \mu^{(d-6)/2}$, where $\mu$ is an arbitrary
momentum scale, $B_d = \Gamma(4-d/2) / (4 \pi)^{d/2}$ is a geometric factor,
and $r_{0c}$ denotes the fluctuation--induced shift of the percolation
threshold,
\begin{equation}
     r_{0c} = \left[
              {u_0^2 c_0 B_d \over (d-4) (6-d)} I^d_{13}(g_0/\sqrt{r_{0c}})
                                             \right]^{2 / (6 - d)} \quad .
\label{4}
\end{equation}
Here we defined $I^d_{mn}(g) = \int_0^1 {x^{m/2 - 1} (1 + x g^2)^{(d-m-n)/2}}
dx $, with the limits $I^d_{mn}(0) = 2/m$, and $\lim_{g \rightarrow \infty}
[g^m I^d_{mn}(g)] = \Gamma(m/2) \Gamma((n-d)/2) / \Gamma((m+n-d)/2)$.

The renormalization group (RG) equation explicitly takes advantage of the scale
invariance of the system near the percolation threshold. More precisely, we
observe that the bare pair correlation functions do not depend on the arbitrary
renormalization scale $\mu$. By introducing Wilson's flow functions (we list
the explicit one-loop results here; the symbol $\vert_0$ indicates that all the
derivatives are to be taken at fixed bare parameters)
\begin{eqnarray}
    &&\zeta_\phi = \mu {\partial \over \partial \mu} \bigg \vert_0 \ln Z_\phi
                  = {v \over 8}
          \left[ 1 - {I^d_{35}(g) \over I^d_{17}(g)} \right] \, , \quad
      \zeta_r = \mu {\partial \over \partial \mu} \bigg \vert_0
                 \ln {r \over r_0 - r_{0c}} = -2 + {3 v \over 8}
               + {v \over 8} {I^d_{35}(g) \over I^d_{17}(g)} \, ,  \label{5}\\
    &&\zeta_c = \mu {\partial \over \partial \mu} \bigg \vert_0
                 \ln {c \over c_0} = - {v g^2 (d-8) \over 16}
                  \left[ 2 {I^d_{37}(g) \over I^d_{17}(g)} -
                     {I^d_{55}(g) \over I^d_{17}(g)} \right] \, , \label{6}\\
    &&\zeta_g = \mu {\partial \over \partial \mu} \bigg \vert_0
                 \ln {g \over g_0} = -1 - \zeta_\phi + \zeta_c + {v \over 4}
          \left[ 1 - {I^d_{35}(g) \over I^d_{17}(g)} \right]  \, , \label{7}\\
    &&\zeta_u = \mu {\partial \over \partial \mu} \bigg \vert_0
                 \ln {u \over u_0} =
                     {d-6 \over 2} + {13 v \over 16} + {3 v \over 16}
                     {I^d_{35}(g) \over I^d_{17}(g)}         \, ,   \label{8}
\end{eqnarray}
one arrives at the following RG equation for the renormalized
two--point vertex function
\begin{equation}
     \left[ \mu {\partial \over \partial \mu} + \! \!
     \sum_{a = \{ r,c,g,u \}} \! \! \! \zeta_a a {\partial \over \partial a} +
         \zeta_\phi \right]
         \Gamma_{11}(\mu,r,c,g,u,{\bf q},\omega) = 0 \quad .
\label{11}
\end{equation}

The RG equation (\ref{11}) is solved by introducing the characteristics
$a(\ell)$, which define the running parameters and couplings. They are given by
the solutions of $\ell {d a(\ell) / d \ell} = \zeta_a(\ell) a(\ell)$, with
$a(1) = a$. Defining the dimensionless vertex function ${\hat \Gamma}_{11}$
according to $\Gamma_{11}(\mu, r, c, g, u, {\bf q}, \omega) = \mu^2 {\hat
\Gamma}_{11} (r, v, {\bf q} / \mu, g \omega / c \mu, \omega^2 / c^2 \mu^2)$,
the solution of Eq.~(\ref{11}) reads
\begin{equation}
     \Gamma_{11}(\mu,r,c,g,u,{\bf q},\omega) =
      \mu^2 \ell^2 e^{ \int_1^\ell \zeta_\phi(\ell') d\ell' / \ell' }
       {\hat \Gamma}_{11} \left( r(\ell), v(\ell), {{\bf q} \over \mu \ell},
                           {g(\ell) \omega \over c(\ell) \mu \ell},
                      {\omega^2 \over c(\ell)^2 \mu^2 \ell^2} \right) \quad .
\label{13}
\end{equation}
Here we have introduced an effective anharmonic coupling $v = u^2 c
I^d_{17}(g)$, which is finite in both limits, $g \rightarrow 0$ and $g
\rightarrow \infty$. The flow of the running coupling $\ell {d v(\ell) / d
\ell} = \beta_v(\ell)$ is given by the corresponding $\beta$ function $\beta_v
= \mu {\partial v / \partial \mu}  \vert_0$. The flow parameter $\ell$ may be
considered as describing the effect of a scaling transformation upon the
system. The theory becomes scale--invariant near a fixed point $v^*$, defined
as a zero of the $\beta$ function, and will yield the correct asymptotic
behavior, if it is infrared--stable.

We now turn to study Eq.~(\ref{13}) in the vicinity of such a fixed point
$v^*$; introducing $\zeta_a^* = \zeta_a(v = v^*)$ we find that $\Gamma_{11}$ is
a generalized homogeneous function
\begin{equation}
  \Gamma_{11}(\mu,r,c,g,u,{\bf q},\omega) \propto \mu^2 \ell^{2 + \zeta_\phi^*}
      {\hat \Gamma}_{11} \left( r \ell^{\zeta_r^*}, v^*, {{\bf q}
                  \over \mu \ell},
                  {g / c \mu  \omega \over \ell^{1 + \zeta_c^* - \zeta_g^*}},
         {\omega^2 / c^2 \mu^2 \over \ell^{2 (1 + \zeta_c^*)}} \right) \quad .
\label{15}
\end{equation}
Using an appropriate matching condition, we can now map the asymptotic theory
with manifest infrared divergences, onto a region in parameter space where the
anharmonic coupling is finite. E.g., with the choice ${\ell = q / \mu}$, we
arrive at the following {\it self--affine} scaling form
\begin{equation}
     \Gamma_{11}(\mu,r,c,g,u,{\bf q},\omega) \propto q^{2 - \eta_\perp}
      {\hat \Gamma}_{11} \left( {r \over (q / \mu)^{1 / \nu_\perp}}, v^*, 1,
                          {g / c \mu \omega \over (q / \mu)^z},
      {\omega^2 / c^2 \mu^2 \over (q / \mu)^{2 z (1 - \Delta)}} \right) \, ,
\label{16}
\end{equation}
where we have defined {\it four} independent critical exponents according to
$\eta_\perp = - \zeta_\phi^*$, $\nu_\perp = - 1 / \zeta_r^*$, $z = 1 +
\zeta_c^* - \zeta_g^*$, and $z \Delta = - \zeta_g^*$. Here, $\eta_\perp$ and
$\nu_\perp$ correspond to the two independent indices familiar from the theory
of static critical phenomena. The ``dynamic'' exponent $z$ is related to the
anisotropic scaling behavior. Finally, $\Delta$ is a positive crossover
exponent describing the transition from isotropic to directed percolation. In
the asymptotic limit of directed percolation, $g \rightarrow \infty$, the
second scaling variable disappears, and the scaling behavior is described by
the {\it three} exponents $\eta_\perp$, $\nu_\perp$, and $z$.

Similarly, with the choice $\ell = \left( g \omega / c \mu \right)^{1 / (1 +
\zeta_c^* - \zeta_g^*)}$ Eq.~(\ref{15}) reads
\begin{equation}
     \Gamma_{11}(\mu,r,c,g,u,{\bf q},\omega)
     \propto \omega^{2 - \eta_\parallel}
      {\hat \Gamma}_{11} \left( {r \over (q / \mu)^{1 / \nu_\parallel}}, v^*,
             {q / \mu \over (g \omega / c \mu)^{1 / z}}, 1,
  {\omega^2 / c^2 \mu^2  \over  (g \omega / c \mu)^{2 (1 - \Delta)}} \right)
                                                            \, ,
\label{18}
\end{equation}
where $2 - \eta_\parallel = (2 - \eta_\perp)/z$ and $\nu_\parallel = z
\nu_\perp$. Moreover, $\ell = r^{-1 / \zeta_r^*}$ leads to
\begin{equation}
     \Gamma_{11}(\mu,r,c,g,u,{\bf q},\omega) \propto r^\gamma
      {\hat \Gamma}_{11} \left( 1, v^*, {q \over \mu} r^{- \nu_\perp},
                 {g \omega \over c \mu} r^{- \nu_\parallel},
 {\omega^2 \over c^2 \mu^2} r^{-2 \nu_\parallel (1 - \Delta)} \right) \quad ,
\label{20}
\end{equation}
where the exponent $\gamma$ is related to $\nu_\perp$ and $\nu_\parallel$ via
$\gamma = \nu_\perp (2 - \eta_\perp) = \nu_\parallel (2 - \eta_\parallel)$. In
the special case of isotropic percolation, the scaling relations simplify
considerably, describing {\it self--similar} scaling with only {\it two}
independent critical exponents $\eta = - \zeta_\phi^*$ and $\nu = -1 /
\zeta_r^*$.

In the two limiting cases of isotropic and directed percolation, we can
explicitly determine the critical indices to one--loop order. For $g_0 = 0$ and
$d < d_c^{\rm I} = 6$ one finds a stable, nontrivial fixed point $v^*_{\rm I} =
4 (6 - d) / 7$, while for $d \geq 6$ the Gaussian fixed point $v^*_{\rm GI} =
0$ is approached. In the opposite case, $g_0 \rightarrow \infty$, the
nontrivial directed fixed point is stable only if $d < d_c^{\rm D} = 5$, while
for $d = D + 1 \geq 5$ the mean--field exponents of the Gaussian fixed point
$v^*_{\rm GD}$ apply. The fixed--point values and the corresponding independent
critical indices are collected in Table I (there the coupling ${\tilde g} = g /
(1+g)$ was defined). The other critical exponents may be inferred from the
scaling relations.

Thus we have demonstrated that both the self--similar and the self--affine
scaling behavior are within the scope of the present theory, at least for
dimensions $d < 5$; for $5 < d \leq 6$ the model is not renormalizable in the
directed limit, and simply characterized by the exponents corresponding to the
Gaussian fixed point $v^*_{\rm GD}$, with logarithmic corrections for $d = 5$.
This again emphasizes the fact that no expansion with respect to a fixed upper
critical dimension can be applied consistently.

The interchange from self--similar to self--affine scaling is most conveniently
described by effective exponents for the pair--correlation function. Using the
zero--loop result for the two--point vertex function
\begin{equation}
     \Gamma_{11} (r,{\bf q},\omega) = \mu^2 \ell^2
       e^{\int_1^\ell \zeta_\phi(\ell') d \ell'/ \ell'}
      \left[ r(\ell) + \left( {q \over \mu \ell} \right)^2 +
                {\omega^2 \over \mu^2 \ell^2 c(\ell)^2} +
               2 i {\omega g (\ell) \over \mu \ell c(\ell)} \right] \; ,
\end{equation}
we specialize to $r = 0$ and ${\bf q} = {\bf 0}$ and define [compare
Eq.~(\ref{18})]
\begin{equation}
     2 - \eta_{\parallel \, \rm eff}(\omega) =
      {d \ln {\sqrt{\mid \Gamma_{11}(0,{\bf 0},\omega) \mid^2 }}
       \over d \ln \omega}       \quad .
\label{23}
\end{equation}
Using $\left \vert \omega^2 / \mu^2 \ell^2 c(\ell)^2 + 2 i \omega g(\ell) / \mu
\ell c(\ell) \right \vert^2 = 1$ yields $2 - \eta_{\parallel \, \rm eff}(\ell)
= [ 2 + \zeta_{\phi}(\ell) ] d \ln \ell / d \ln \omega$. Similarly, in the case
$r = \omega = 0$ [see Eq.~(\ref{16})], let
\begin{equation}
     2 - \eta_{\perp \, {\rm eff}}(q) = { d \ln \Gamma_{11}(0,{\bf q},0)
      \over d \ln q} \quad ,
\label{24}
\end{equation}
which reduces to $2 - \eta_{\perp \, {\rm eff}} = 2 + \zeta_\phi(\ell)$, if $(q
/ \mu \ell)^2 = 1$ is inserted. Finally, considering ${\bf q} = {\bf 0}$ and
$\omega = 0$ we introduce [compare Eq.~(\ref{20})]
\begin{equation}
    \gamma_{\rm eff}(r) = {d \ln \Gamma_{11}(r,0,0) \over d \ln r} \quad ,
\label{25}
\end{equation}
and choosing the matching condition $r(\ell) = 1$ we find $\gamma_{\rm
eff}(\ell) = - [2 + \zeta_\phi(\ell)] / \zeta_r(\ell)$.

The flow of the effective exponents $\eta_{\parallel \, \rm eff}(\ell)$,
$\eta_{\perp \, \rm eff}(\ell)$, and $\gamma_{\rm eff}(\ell)$ in $d = D + 1 =
3$ dimensions (with $\mu = 1$) is depicted in Fig.~1, with the initial value
for the coupling $v(1) = v^*_{\rm I}$ of the isotropic scaling fixed point. The
dependence on the anisotropy scale $g$ was eliminated by plotting versus the
scaling variable $\ln g(\ell)$; the graphs corresponding to different initial
values $g(1)$ then all collaps onto one master curve. The most important
conclusion to be drawn from Fig.~1 is that the anisotropy scale, at which the
crossover occurs, considerably differs for the effective exponents defined
above ! $\eta_{\parallel \, \rm eff}$ starts to cross over from the isotropic
to the directed fixed point value already at $\ln g(\ell_{\rm cross}) \approx
-0.8$, whereas $\gamma_{\rm eff}$ shows this crossover at $\ln g(\ell_{\rm
cross}) \approx -0.2$, and $\eta_{\perp \, \rm eff}$ only at $\ln g(\ell_{\rm
cross}) \approx +0.8$. Note that the sizeable change of $\eta_{\parallel \, \rm
eff}$ is already apparent at mean-field level, where it acquires the values $0$
and $1$ in the isotropic and directed limit, respectively. However, a crossover
of the exponents $\eta_{\perp \, \rm eff}$ and $\gamma_{\rm eff}$ requires the
$\zeta$ functions at least on the one--loop level.

We remark that a calculation of the above crossover features has not
been possible up to this present work. Of course, the exponents for the limits
of both isotropic and directed percolation have been determined to a much
higher accuracy than is provided in our one--loop approximation
\cite{Ess80,Kin83,Car80,Ben84}. However, our aim was rather to describe the
crossover features, and we believe that the crossover loci should not be
affected too severely by, say, higher orders of perturbation theory. The
expectation that our {\it ``renormalized mean--field theory''}, accompanied
with the one--loop results for the flowing parameters, is a reasonably good
approximation, is based on the experience that amplitude functions are usually
smooth and enter the results less sensitively than the exponent functions. In
fact, our method was designed to incorporate the entire crossover behavior into
the Wilson flow functions. The proposed approach to crossover phenomena, which
is an extension of Amit and Goldschmidt's ``generalized minimal subtraction''
scheme \cite{Ami78}, should thus be applicable to a large variety of crossover
phenomena. It would therefore be of considerable interest to compare our
predictions concerning the crossover scales of the different effective
exponents with the outcome of precise computer simulations and/or experiments
in order to estimate the quality of our approximations also in different
situations, where numerical simulations are either very cumbersome or not
feasible at all.


\noindent {\bf Acknowledgement:} {\small This work has been supported by the
Deutsche Forschungsgemeinschaft (DFG) under Contracts No. Fr. 850/2-1,2,
Ta. 177/1-1, and Schw. 348/4-2.}




\mediumtext
\begin{table}
\setdec 0.00
\caption{Fixed point values for the couplings $v(\ell)$ and ${\tilde g}(\ell)$
at the isotropic (I), Gaussian isotropic (GI), directed (D), and Gaussian
directed (GD) fixed points, respectively, and the corresponding values of the
four independent critical exponents.}
\medskip

\begin{tabular}{lcccccc}
{\rm Fixed Point}    &  $v^*$  &  ${\tilde g}^*$  &  $\eta_\perp$ &
$\nu_\perp^{-1}$  &  $z$  & $\Delta$      \\
\tableline
{\rm GI}     &  $0$ & $0$ & $0$ & ${2}$ & $2$ & ${1 \over 2}$ \\
{\rm I}      & ${4 \over 7} (6-d)$ & $0$ & $-{6-d \over 21}$ & $2 - {5 (6-d)
\over 21}$ & $2 - {6-d \over 21}$ & ${1 - (6-d)/21 \over 2 - (6-d)/21}$  \\
{\rm GD}     &  $0$ & $1$ & $0$ & ${2}$ & $2$ & ${1 \over 2}$ \\
{\rm D}      &  ${2 \over 3} (5-d)$ & $1$ & $- {5-d \over 12}$ & $2 - {5-d
\over 4}$ & $2 - {5-d \over 12}$ & ${1-(5-d)/6 \over 2 - (5-d)/12}$ \\
\end{tabular}
\label{table1}
\end{table}


\begin{figure}
\epsfxsize=5.0truein
\epsffile{figure_el.eps.bb}
\medskip
\caption{Effective exponents $\eta_{\parallel \, {\rm eff}}(\ell)$ (solid),
$\eta_{\perp \, {\rm eff}}(\ell)$ (dashed), and $\gamma_{\rm eff}(\ell)$
(dotted) for the connectivity as functions of the running anisotropy parameter
$g(\ell)$ for a fixed initial value of the coupling constant $v(1) = v^*_{\rm
I}$.}
\label{fig1}
\end{figure}

\end{document}